# Nanometre scale monitoring of the quantum confined stark effect and emission efficiency droop in multiple GaN/AlN quantum disks in nanowires

L. F. Zagonel[1*], L. H. G. Tizei[2], G. Z. Vitiello[1], G. Jacopin[3], L. Rigutti[4], M. Tchernycheva[5], F. H. Julien[5], R. Songmuang[6], T. Ostasevicius[7], F. de la Peña[7], C. Ducati[7], P. A Midgley[7], M. Kociak[2**]

1- Instituto de Física "Gleb Wataghin" Universidade Estadual de Campinas – Unicamp, 13083-859, Campinas, São Paulo, Brasil
2- Laboratoire de Physique des Solides, CNRS, Univ. Paris-Sud, Université Paris-Saclay, 91405 Orsay Cedex, France
3- Institute of Physics, Ecole Polytechnique Fédérale de Lausanne, 1015 Lausanne, Switzerland
4- Groupe de Physique des Matériaux, UMR CNRS 6634, Université et INSA de Rouen, Normandie University, 76800 St Etienne du Rouvray, France
5- Institut d'Electronique Fondamentale, University Paris Sud, University Paris Saclay, Orsay 91405, France
6- CEA-CNRS group 'Nanophysique et Semiconducteurs', Institute Neel, Grenoble 38054, France
7- Department of Materials Science and Metallurgy, University of Cambridge, Cambridge CB3 0FS, UK

## ABSTRACT

We report on a detailed study of the intensity dependent optical properties of individual GaN/AlN Quantum Disks (QDisks) embedded into GaN nanowires (NW). The structural and optical properties of the QDisks were probed by high spatial resolution cathodoluminescence (CL) in a scanning transmission electron microscope (STEM). By exciting the QDisks with a nanometric electron beam at currents spanning over 3 orders of magnitude, strong non-linearities (energy shifts) in the light emission are observed. In particular, we find that the amount of energy shift depends on the emission rate and on the QDisk morphology (size, position along the NW and shell thickness). For thick QDisks (>4nm), the QDisk emission energy is observed to blue-shift with the increase of the emission intensity. This is interpreted as a consequence of the increase of carriers density excited by the incident electron beam inside the QDisks, which screens the internal electric field and thus reduces the quantum confined Stark effect (QCSE) present in these QDisks. For thinner QDisks (<3 nm), the blue-shift is almost absent in agreement with the negligible QCSE at such sizes. For QDisks of intermediate sizes there exists a current threshold above which the energy shifts, marking the transition from unscreened to partially screened QCSE. From the threshold value we estimate the lifetime in the unscreened regime. These observations suggest that, counterintuitively, electrons of high energy can behave ultimately as single electron-hole pair generators. In addition, when we increase the current from 1 pA to 10 pA the light emission efficiency drops by more than one order of magnitude. This reduction of the emission efficiency is a manifestation of the 'efficiency droop' as observed in nitride-based 2D light emitting diodes, a phenomenon tentatively attributed to the Auger effect.

* zagonel@ifi.unicamp.br

** mathieu.kociak@u-psud.fr

# I. Introduction

Nanoscale structuring of materials is a promising route to engineer material properties and achieve new functionalities unthinkable in conventional devices. In particular, semiconductor nanowires (NWs), *i.e.* high aspect ratio nano-crystals with the lateral dimension in the range of few to tens of nanometres, have shown their potential for a new generation of optoelectronic devices.[1,2,3] Indeed, some very interesting examples of nanowire-based devices have already been demonstrated in the laboratory. Many of them contain quantum-confined heterostructures made up of different materials in order to achieve controlled-by-design electronic and optical properties.[4,5,6] III-nitride semiconductor nanowires for light emitting devices in the visible-UV range have already been demonstrated [7,8]. However, a thorough understanding of such complex system's fine properties and subtleties is required to engineer and control device performance.

One current issue is to eliminate the droop of emission efficiency occurring at large carrier injection inside the optically active areas in both thin films [9,10,11,12] and NWs [13,14]. The carrier density also affects the wavelength of the emitted light.[15] Indeed, the presence of high internal electric fields in polar III-N heterostructures induces a strong band distortion in quantum-confined structures [16,17]. This results in a large redshift of the transition energy with respect to a flat-band case (so-called Quantum Confined Stark Effect (QCSE)) and in a spatial separation of the electrons and holes reducing the emission rate [18,19,20,21,22,23,24,25]. When the charge carrier density increases, non-linearities are expected [19,26,27,28,29,30,31] as charge carriers screen the internal field and thus reduce the redshift and increase the emission rate. On the other hand, the increase in charge carrier density leads to the upturn of high order phenomena such as the Auger effect (the non-radiative recombination of an electron-hole pair with an energy transfer to a third charge in their vicinity), which are deleterious for the emission properties [12].

Therefore, manipulating the carrier density and determining its influence on the emission wavelength and on the quantum efficiency are key prerequisites toward the understanding of nitride devices based on quantum confined electronic states. A large number of studies [11,20,24,32,33,34] have already addressed this problem in the case of thin films. When the active material in nitride devices is replaced by 3D NWs, new characterization and understanding challenges appear [35,36], which today remain poorly addressed.[37]

In the case of closely packed quantum emitters [14,38,39], the characterization of optical properties requires high spatial resolution in order to get both the morphology and optical activity at relevant spatial and spectral scales. Such requirements are better met by Transmission Electron Microscope operated in scanning mode (STEM) associated with high performance cathodoluminescence detectors (CL). [40,41] For a review, see ref. [42]. In this technique, a finely focused electron beam (not subject to light-optics diffraction limit) excites small regions of a sample (as small as 1 nm$^2$, or even less) and the emitted photons are collected and then analysed by an optical spectrometer. Moreover, cathodoluminescence can be used in the so-called spectrum-imaging (SPIM) mode (also called hyperspectral imaging) [43] where full spectra are acquired on a whole region of the sample for each electron beam position by scanning the electron beam. Simultaneously, a structural or morphological image, typically a High Angle Annular Dark

Field (HAADF), can be acquired. This operation mode provides very accurate simultaneous spatial and spectral information, giving clear correlation between light emission and the position of its excitation, a key to relate optical properties to other local properties as morphology, composition, strain, etc. Indeed, CL techniques have been successfully applied to gain understanding of different systems, including III-V heterostructures [25, 40,44,45,46,47,35,48,49,50,51], diamond [52,53,54], plasmonics nanostructures [55,56,57,58] among others.

Here, we have applied CL-STEM to study the influence of charge carrier density on the optical response of individual GaN quantum disks (QDisks) confined by AlN barriers in NWs. A careful and systematic analysis of the emission energy and intensity of many QDisks at controlled excitation currents makes possible to observe correlations between the energy shifts, emission rates and QDisks morphology which are interpreted in terms of the screening of internal electric fields and reduction of the QCSE. The GaN/AlN heterostructured nanowires were chosen since they represent a model system with a strong internal field and band gap engineering (similarly to InGaN layers), which at the same time has low defect density and sharp concentration gradients.[59] The analysis of the emission efficiency (i.e. the emission rate normalized to the incident electron beam current) demonstrates the presence of an emission efficiency droop starting at ~1 pA of electron beam current (despite the use of the STEM electron probe producing a small number of carriers inside the QDisk). The use of the CL technique for local carrier generation in the AlN/GaN system presenting strong carrier localization allows minimizing charge transport effects. The observed quantum efficiency droop is tentatively attributed to Auger effects.

## II. Sample Experimental Methods

GaN NWs containing 20 GaN/AlN QDisks have been grown at 790 °C by catalyst-free Plasma-Assisted Molecular Beam Epitaxy (PA-MBE) on Si (111) substrate. More details about the growth can be found in [60]. A typical nanowire consists of a ~0.5 μm GaN base part, followed by a series of GaN/AlN QDisks with a nominal thickness of 3 nm and of a ~0.5 μm GaN cap part. As seen from Fig. 1, the QDisk thickness increases along the growth direction from ~1 nm to ~4.5 nm, while the AlN barriers have thicknesses from 2.6 to 3.6 nm independently of the growth order. The QDisks are formed by switching from Ga to Al flux without growth interruption. Due to the AlN lateral growth, the QDisk region and the GaN base are surrounded by an external AlN shell, while the GaN grown latter remains uncovered [61]. For the CL-STEM studies we have deliberately selected large diameter NWs in which thick QDisks (>4 nm) are likely to be present, in contrast to our former studies of thin QDisks [40]. The NWs were dispersed on a standard thin carbon film on a copper grid. For CL-STEM SPIM experiments, they were analysed in a VG HB501 STEM working at 60 keV using an in-house built nanoCL system, which has been optimized for high speed and high spatial resolution spectral imaging [40]. A cold finger at the sample holder keeps the sample at about 150 K. The calibration procedure for emission rate estimation is given in the Annex. For high resolution HAADF (HR-HAADF), the samples were analysed in a NION USTEM 200 operated at 200 keV.

In order to explore different regimes of the light emission rate, the incident electron beam current was set typically in the range between 0.1 pA and 600 pA. The electron beam current was monitored during the experiments using an Electron Energy Loss spectrometer as a Faraday cup and by measuring the current with an electrometer. In addition, various acquisition dwell times per spectra were used, ranging from approximately 20 ms to 10 s.

Differently from previous studies [40, 41] where all or most QDisks were studied, here we performed the in depth analysis only on selected QDisks with light emission spectrally and spatially isolated from others due to the much similar emission energy for larger QDisks and so that energy shift could be tracked. We present a representative compilation of 16 SPIMs on 4 NWs, corresponding to a total of 80 QDisks. The NWs are labelled (a), (b), (c) and (d) and the QDisks are indexed after their growth order in each NW. Each CL SPIM contains from 1k (250 by 4 positions in the sample plane) to 30k (300 by 100) spectra.

In order to estimate the emission energy and intensity of CL signal we fit a simple model to the spectra using non-linear least squares weighted to account for Poisson noise. The model consists of a Lorentzian function and a background that is flat in the spectral region of interest. We perform multi-dimensional curve fitting on the SPIMs using the SAMFire algorithm as found in the development version of HyperSpy.[62] SAMFire automatically estimates the starting parameters at each spectra to avoid falling in local minima, easing the task of performing curve fitting in multi-dimensional datasets.

# III. RESULTS AND DISCUSSION

### III.A EMISSION ENERGY TO EMISSION RATE RELATION

Fig. 1(a) shows the morphology of a typical NW, labelled NW(a), with its QDisks. This image was acquired simultaneously with a SPIM. The results of the SPIM are condensed in Fig. 1(b), which is obtained by adding all energy-filtered images from the SPIM after colouring them according to a colour scale associated to the emission energy. No drift correction has been applied to the data resulting in a distortion of the image and SPIM (the total acquisition time was about 8 minutes) with respect to the actual size of the object. The NW growth direction is from right to left. The full CL SPIM is given in the Supplementary Materials as a video [63], displaying the spatial distribution and intensity of light emission for a wide spatial region of the sample within a broad spectral range. Some slices from this video are shown in Fig. 2. It shows that CL intensity maxima are centred on different QDisks, depending on the emission energy. Roughly, the smaller the QDisk the higher will be its emission energy, as expected from quantum confinement. The effect of the barriers is also observed as the spatial width (at 1/e) of the light emission intensity from a single QDisk is about 10 nm (i.e.: from the centre of the QDisk to about 5 nm away in both directions along the NW growth axis). This is very small when compared to the carrier diffusion in bulk GaN or AlN [54,40] or GaN nanowires, [64] revealing the strong carrier trapping capability of the QDisks. However in the present case, contrary to [40], a clear broadening in the spatial distribution of the detected signal arises at lower energy – a given QDisk appears broader on filtered map of smaller energy, indicating a link between excitation position and emission energy. Attribution of spectral features to individual QDisks is however possible, as detailed in the Annex.

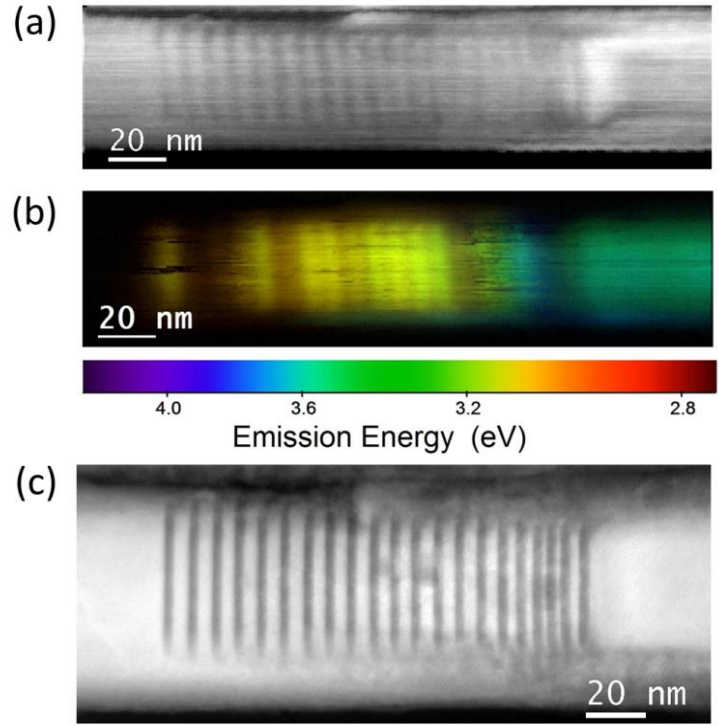

*Fig. 1: (a) and (c) are HAADF images of NW(a) containing 20 QDisks of GaN within AlN barriers. The image shown in (a) was taken simultaneously with the SPIM and shows some vertical compression due to sample drift. The Fig. in (b) shows a false colour image obtained by colouring each energy filtered image according to the colour scale shown and summing them. The image shown in (c) shows the same NW in HR-HAADF-STEM. GaN appears as light grey while AlN appears as dark grey.*

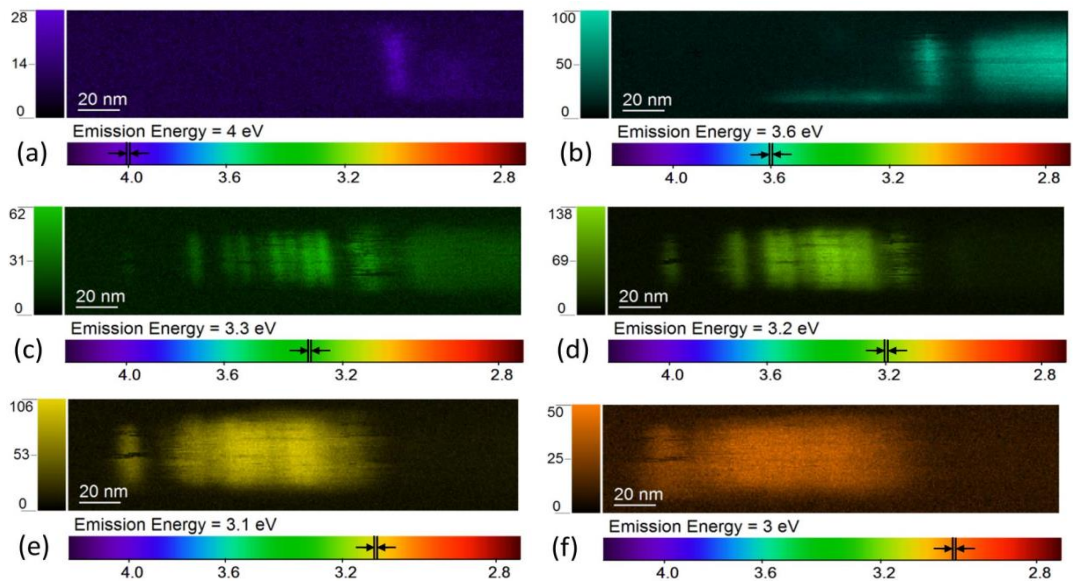

*Fig. 2: These images display 6 slices of the spectrum image from Fig. 1(b) showing the spatial excitation maps for each indicated energy. In each CL image, a different (false) colour helps indicating the energy according to the colour scale on the bottom. The intensity is indicated by the gradient scale on the left-hand side, which gives also the intensity range in each image. The NW growth direction is from right to left.*

The relationship between the emission energy and the excitation position is made clearer in Fig. 3 were the spatial, spectral and intensity behaviour of two different QDisks are emphasized. On Fig. 3(a-d), we focus on QDisk No. 20 (NW(a)) shown above, which has a thickness of 4.7 nm and an emission spatially and spectrally well separated from that of other QDisks. The emission intensity decreases as the electron beam moves away from the QDisk. From averaged spectra around regions within a certain intensity range (Fig. 3(c-d)) it is clear that lower intensity is linked with lower emission energy. Moreover, at high excitation close to the QDisk centre the spectrum is peaked at higher energy and contains a shoulder, indicating that a range of emission energies are present under the same excitation conditions, in agreement with the stochastic nature of the excitation process.

The connection between energy shift and intensity change can be confirmed by following spectral changes of a given QDisk as a function of incident beam current for a fixed beam position, as shown for QDisk No. 10 of NW(b) (Fig. 3(e)). The variation is not linear with emission intensity. Initially, no shift is observed at small intensity (Fig. 3 (f)) up to roughly $10^3$ count/s. At higher count rates a continuous blue-shift is seen, what is true for most QDisks emitting at energies below GaN bulk band gap, as discussed in more detail later.

As the blue-shift occurs due to changes in the excitation position and/or the excitation current, one is tempted to link it to the carrier density created at the QDisk. Thus, the emission rate will depend either on the probe distance to the QDisk at constant incident current or on the incident current for a fixed probe position. A low electron probe current on the centre of the QDisk can be equivalent to a high current probe a few nanometres away from the QDisk centre if both situations yield the same light emission rate. Therefore, we suppose that on average the emission intensity will depend on the density of carriers excited in the QDisk, and use it to follow the energy shifts.

In this view, the blue-shift can be interpreted as due to the screening of the internal electric field and the reduction of the QCSE, as already being suggested by Jahn and co-workers [45] on an indistinguishable ensemble on quantum wells or on basal stacking faults or zinc-blende segments in GaN [64]. Blue-shifts related to field screening have also been seen in time-resolved, non-spatially resolved CL [25]. Indeed, the QCSE red-shifts the emission energy to values below the GaN band gap. Screening of the electric field reduces the QCSE, reducing the red-shift. [19,26,65,66,67]. The absence of new emission peaks or significant broadening in Fig. 3(e) supports our interpretation.

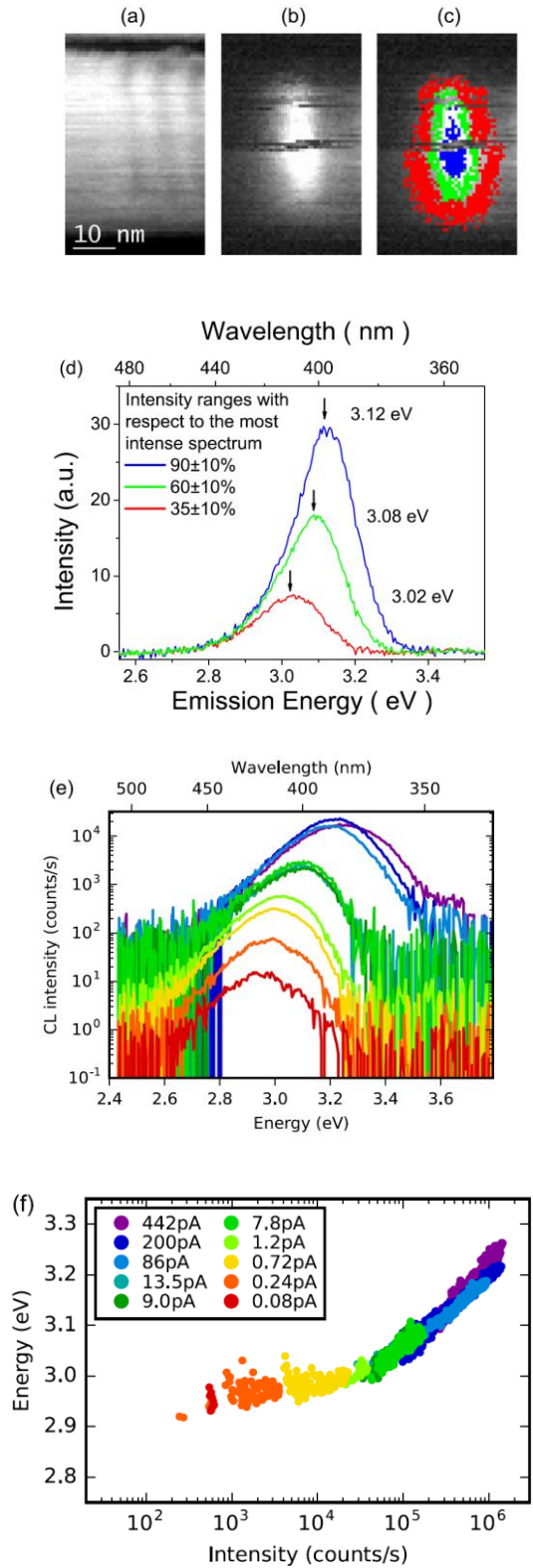

*Fig. 3: (a) Region of the NW(a) containing two QDisks. (b) Panchromatic CL map shows light emission only from the last QDisk (left-most), No. 20, with a thickness of 4.7 nm. Regions of the map with spectra within similar emission intensity ranges are marked in colours in (c) and were averaged to yield spectra representative of each intensity range shown in (d). (e) Typical spectra from QDisk No. 10, with 4.3nm, of NW(b) obtained from geometrical centre of the QDisk in each SPIM. 10 SPIMs are considered, each acquired with a different electron beam current. The much higher range of luminescence intensity as compared to (a) explains the need for a logarithm representation. (f) Emission energy as a function of the emission intensity for QDisk No. 10 considering all 10 SPIMs (each colour indicates a different SPIM with a given electron beam current).*

In order to check that the above-mentioned heuristic picture is valid, we performed some 1D simulations of the luminescence energy of QWs as a simplified model of the present QDisks. The simulations are based on a one-dimensional Schrodinger Poisson solver using the effective mass approximation [24] (material parameters of [24]), taking into account the built-in internal electric field (4 MV/cm [40]), electron-hole carrier density and exciton binding energy [68]. Band diagrams for typical results are shown in Fig. 4. For the sake of the demonstration, the electro-generated carrier density out of equilibrium has been chosen to be $10^{13}$ cm$^{-2}$; a value inducing detectable energy shifts. At equilibrium and subject to a strong internal field (top panels in Fig. 4), the QW conduction and valence (black) bands bend reducing the transition energy. Also, the squared absolute value of the electron (blue) and hole (red) wavefunctions are peaked at different spatial positions, leading to a smaller overlap and a smaller radiative rate. The effect is greater for thicker QWs (right panels on Fig. 4) than for thinner QWs (left panels). When the QWs are filled with carriers, the internal field is screened (bottom panels in Fig. 4), the bending of the bands is reduced and the transition energy becomes larger than in the case of the unscreened QCSE, while the radiative rate increases. It is worth to note that when comparing the 1D simulations to the present experiments, we are considering quantum wells and not quantum disks. The rather large diameter of the disks makes such comparison suitable but one must bear in mind that strain effects, which are present in this system, are not taken into account. In any case, the general tendency observed in the experiments is well reproduced here: at low electrogenerated charge carrier density the transition emission is close to the equilibrium one, while it blue-shifts at high charge carrier density.

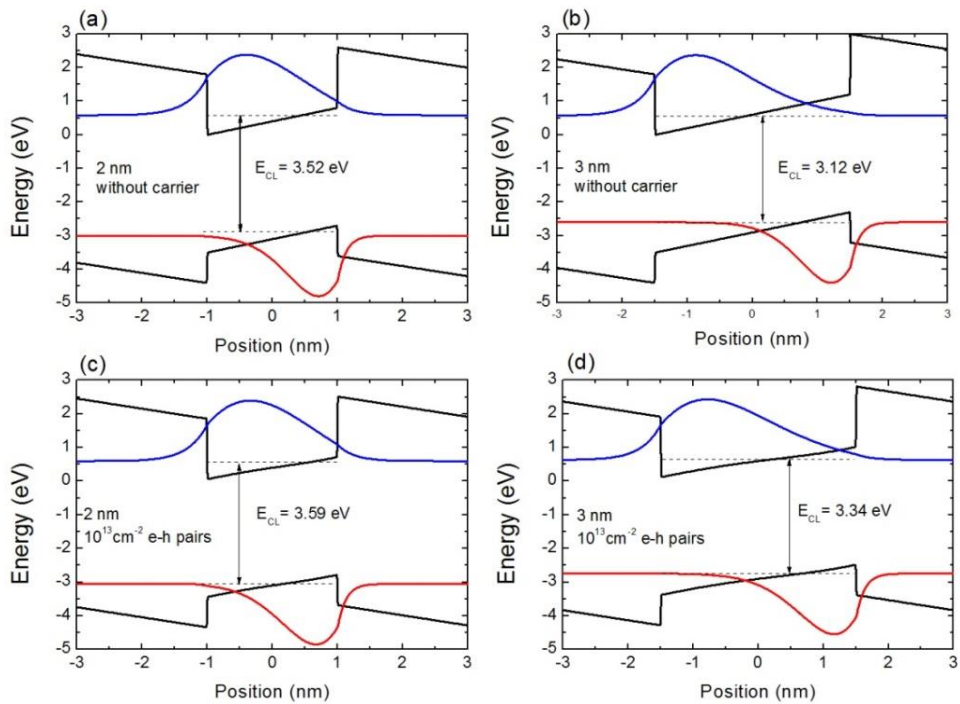

*Fig. 4: 1D simulation of the screened and unscreened QCSE for different thicknesses. The black lines represent the band structure of the wells. The blue (resp. red) curve represents the squared electron (resp. hole) wavefunction of the lowest QW confined level. The dotted lines represent their energy position.*

Therefore, the emission energy can be used as an indicator to monitor the effects of the carrier concentration in a given QDisk. Fig. 5 (a) shows the estimated CL energy as a function of the carrier density for different QW thicknesses. It is worth noting that no significant energy shift is observed below $10^{11}$ cm$^{-2}$ whatever the thickness of the QW. Experimentally, in our datasets, shifts of roughly 0.05 eV are readily spotted, giving us a lower limit of the carrier density inside the QDisk, depending on the QDisk size. Similar trends are observed for the electron-hole overlap (Fig. 5 (b)). The electron-hole overlap increases for increasing carrier density thanks to the screening of the electric field.

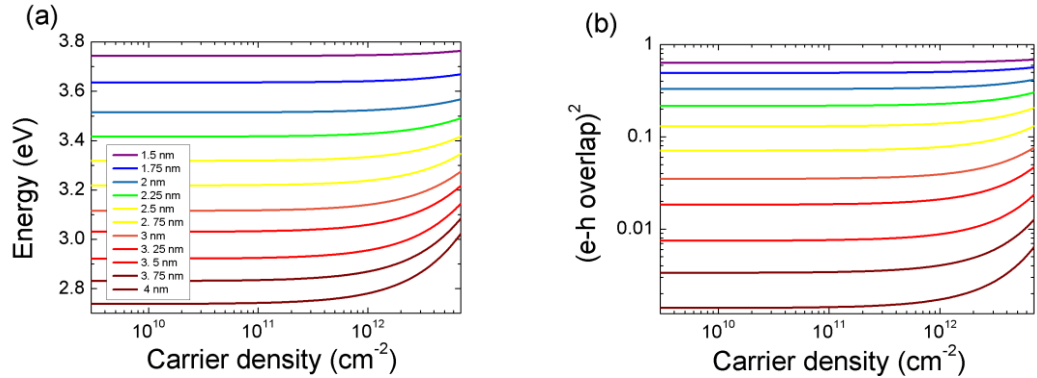

*Fig. 5: a) Simulated CL emission energy as a function of the carrier density in a QW for different QW thicknesses (from 1.5 nm (violet) to 4 nm (red)). (b) Simulated electron-hole wavefunction overlap as a function of the carrier density in a QW for different QW thickness (from 1.5 nm (violet) to 4 nm (red)).*

As it is often the case for the electro or photo-generation of carriers, the carrier density in the QDisks is a hidden variable in our experiments. Indeed, the carrier density in each QDisk depends on the number of electron-hole pairs generated by the electron beam, and also on the radiative and non-radiative lifetimes in each QDisk. In addition, both radiative and non-radiative lifetimes are carrier density dependent, which makes any estimation extremely difficult.

Still, a more general but qualitative comparison between experimental and simulation data can be drawn as exemplified on Fig. 6, where we have plotted the emission energy as a function of the CL intensity (experimental, Fig. 6 (a)) or the product of the electron-hole squared wavefunction overlap by the carrier density (simulation, Fig. 6 (b)) for different thicknesses. This latter comparison is justifiable because CL intensity depends only on the radiative rate and the carrier density $I_{PL} \propto n/\tau_{rad}(n)$, where n is the carrier density and $\tau_{rad}$ is the radiative lifetime) and because the recombination rate is proportional to the squared electron-hole overlap [69].

In Fig. 6 (a), it is possible to note that for thin QDisks, as NW(c) QDisk No. 2 of 2.7 nm, the energy remains the same regardless of the emission intensity. In such small QDisks, the electric field has only a weak influence on the emission energy and therefore the energy remains insensitive to the possible screening of the internal field. In thicker QDisks, starting from about 3 nm, the emission energy is observed to increase with intensity. Moreover, thicker QDisks (5-6 nm) exhibit higher slopes than thinner QDisks (Fig. 6 (a)).

All curves show a clear trend that links the QDisk size to its optical properties under various excitation rates, and that can be qualitatively correlated to the simulations result in Fig. 6 (b).

As stated before the emission rate versus energy relation does not depend directly upon the nominal electron beam current but on the total density of carriers created. This is visually proven by the continuous superposition of different datasets acquired with different beam currents in Fig. 6 (a), which also indicates the absence of cumulative effects or beam damage.

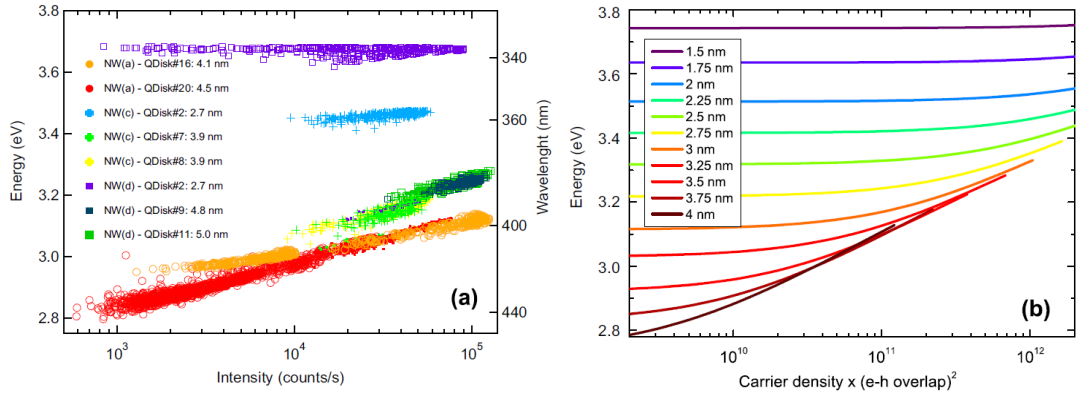

*Fig. 6: (a) Emission energy versus emission intensity extracted from a large number of spectra. Each point in the plot corresponds to a single spectrum. The thickness of each QDisk is indicated in the legend. (b) Simulated CL energy as a function of carrier density multiplied by the electron-hole wavefunction overlap for different QW thicknesses (from 1.5 nm (violet) to 4 nm (red)).*

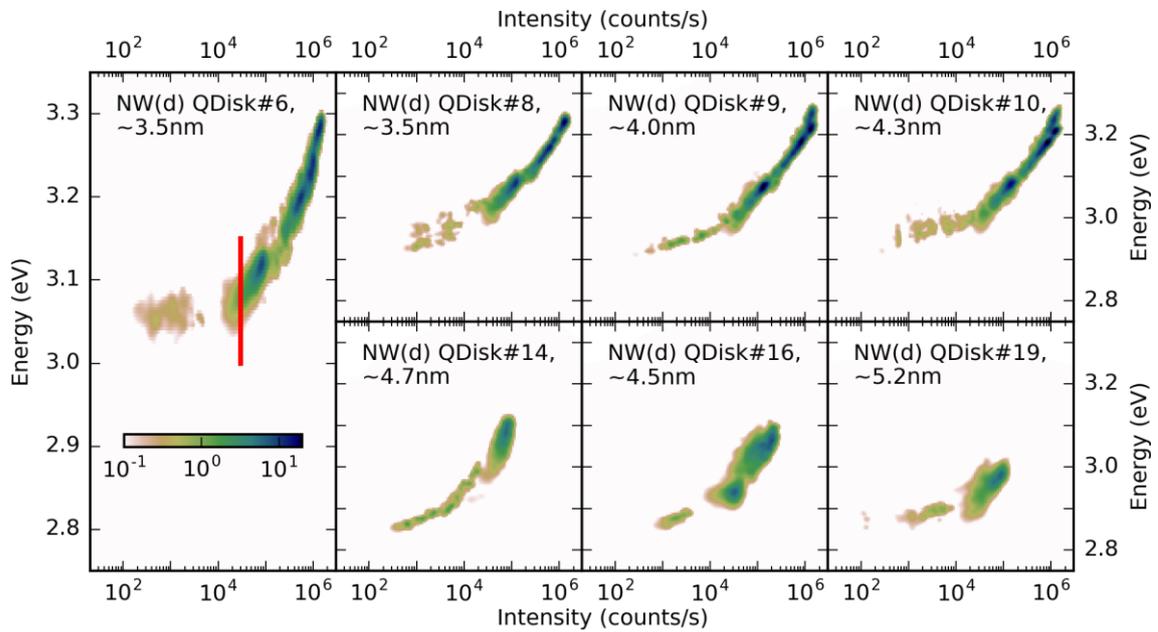

*Fig. 7: Bivariate histogram of emission energy as function of the emission intensity for seven different QDisks on NW(b), regardless of the electron beam current. Each data point has been spread according to its standard deviation. The sum of the intensity of the images equals the total number of spectra considered, 15790. The thickness and QDisk position on the NW are indicated in each panel. While QDisks No. 8 and No. 10 present similar features as discussed above, QDisk No. 6 shows a clear transition from a constant energy to the screening of the electric field responsible for the QCSE (indicated by red lines).*

In Fig. 7 the emission energies vs. emission intensity bivariate histograms are shown for seven QDisks with a higher dynamic range. In this Fig., the data are shown regardless of the SPIM or the electron probe current that was used. By doing this, we indicate the redundancy of this data and its statistical significance. The number of spectra considered in Fig. 7 is 15790. As expected, at high intensity the data have higher density as the datasets have higher number of meaningful spectra.

The higher dynamic range allows the observation of new features not seen in Fig. 6 (a). First, QDisk No. 6 shows a clear transition from constant emission energy, below $\sim 3 \cdot 10^4$ counts/s, to energy shift due to partial screening of the internal electric. This transition marks the excitation rate necessary to create more carriers inside the QDisk than the total recombination rate without internal electric field screening can recombine. The electron beam current at the transition is ~1.2 pA when it is incident at the centre of the QDisk. Considering, as discussed below, that each electron (from the electron beam) creates at most a few electron-hole pairs, [70] we can estimate that each electron creates about one electron-hole pair on average, as some do not suffer inelastic interaction. In this scenario, 1.2 pA corresponds to an excitation of $\sim 7 \cdot 10^9$ carriers per second. Since the QDisk is very close to charge accumulation (and hence to energy shift) the total recombination llifetime is the inverse of the excitation rate, $\sim 1 \cdot 10^2$ ns, in agreement to literature values obtained by time resolved photoluminescence (PL) for this transition energy (ignoring the non-radiative recombination). [19]

Another feature observed in Fig. 7, mainly for QDisks No. 9 and No. 14, is that for the highest observed intensities the slope becomes nearly vertical. Knowing that changes in emission energy are directly linked to changes in carrier density inside each QDisk, the vertical slope means a change in carrier density that does not show up as an intensity increase. Therefore, when the carrier concentration has attained a given value, increasing it (as observed by the energy increase) will not lead to a rise in emission intensity. This phenomenon is associated to a decrease in the conversion efficiency of carriers to photons at high carrier concentration, an effect known as 'efficiency droop'.

For two other QDisks (QDisks No. 16 and No. 19), located close to the end of the heterostructure, the energy variations are much less pronounced. Their maximum emission rate is one order of magnitude smaller than for the most of other QDisks, although the emission has been measured in exactly the same conditions. We attribute this behaviour to the presence of a higher number of non-radiative paths in these QDisks, possibly the recombination at the surface. Indeed, if the non-radiative lifetime is shorter than the radiative one, one expects to observe a lower emission rate and a weaker screening of the internal electric field due to lower carrier density for a given excitation rate. A weaker emission rate and a possibly higher sensitivity to electron irradiation damages are more common for QDisks at the end of the heterostructure possibly due to a thinner AlN shell in this place.

Finally, we consider the variation of the FWHM as a function of intensity and the variation of the emission energy as function of electron beam current (Fig. 8 and Fig. 9). For the wide emission intensity range covering 4 orders of magnitude, the FWHM of most QDisks did not change significantly. However, for QDisks No. 14 and No. 16 some systematic variation is observed, from 0.15 eV to 0.25 eV. For most QDisk the emission energy always increases with the electron beam current, never reaching saturation. These two observations rule out the possibility that the blue-shift of the emission energy or emission efficiency droop as a function of charge carrier density are due to band filling, which would have been a possible explanation otherwise [71].

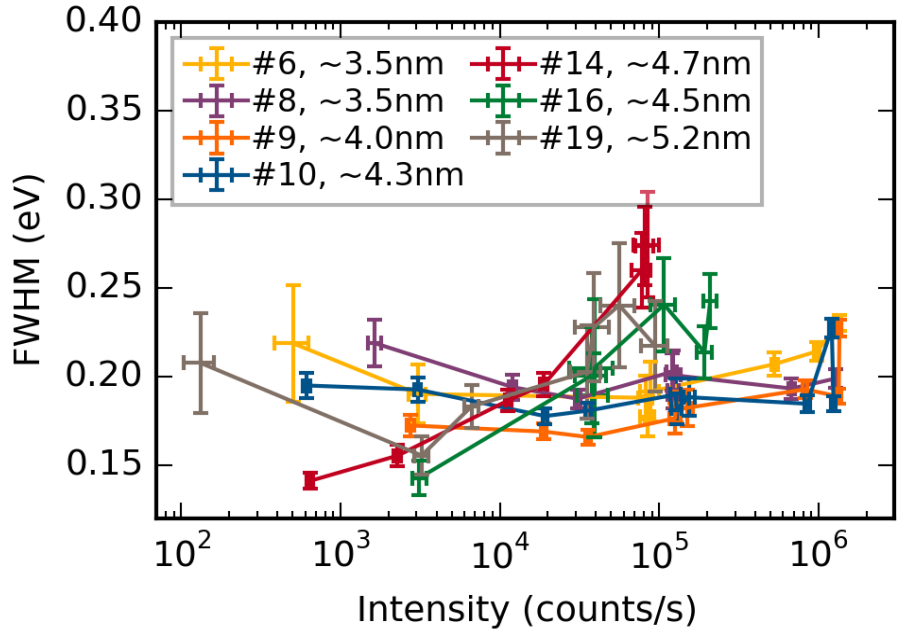

*Fig. 8: Variation of the Full Width at Half Maximum (FWHM) of the light emission spectra for the for QDisks considered in Fig. 7, from NW(b).*

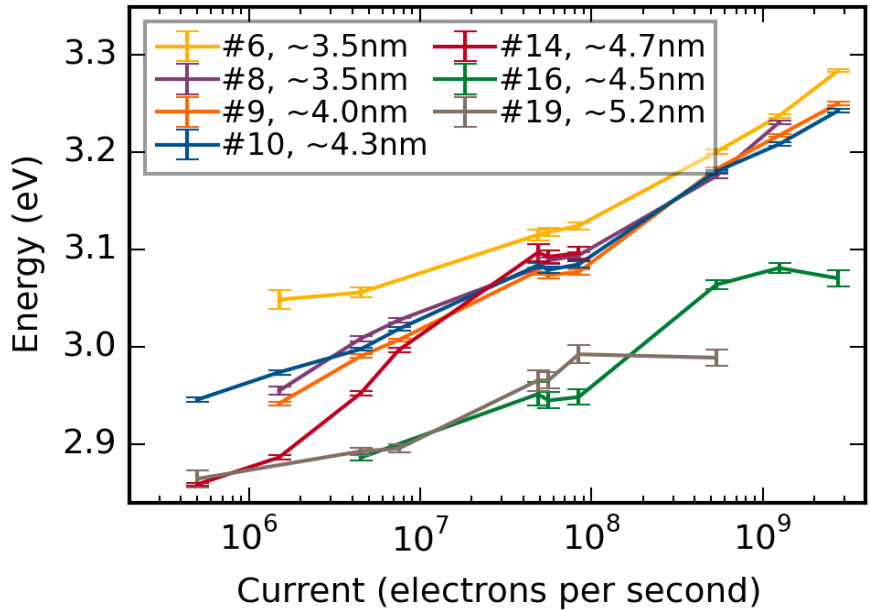

*Fig. 9: Energy shift as function of the electron beam current for the QDisks considered in Figure 7, from NW(b). The thickness of each QDisk is indicated in the legend.*

## III.B Efficiency droop

So far we have considered the emission intensity and energy, regardless of the electron beam current, since the excitation probability depends on the beam distance and position with respect to the QDisk. However, it is important for optoelectronic applications to know the external quantum efficiency (EQE), i.e.: the dependence of the emission rate on the excitation rate. To measure this dependence, we have monitored the emission rate of several QDisks for various electron beam currents when the beam was hitting the centre of the QDisk of interest. To be more specific, we have extracted from each SPIM and for each QDisk, only its highest emission intensity region. For example, this corresponds in Fig. 3 (c) to 9 pixels in the centre of the blue-marked area of the QDisk. Therefore, for this data charge diffusion does not play a role. Under these conditions, the charge carrier density increases with the excitation current (as indicated by Fig. 9).

Fig. 10 presents the ratio of photons emitted per incident electron as a function of the electron beam current, representing the emission efficiency for each excitation rate. For this plot alone, an estimated calibration of counts to absolute photons was used, as shown in the Annex. For low excitation rates from 7 to 100 electrons are necessary for generating a photon emission, depending on the QDisk. This quantitative analysis is in rough agreement with the interaction probability that can be determined by computing the electron mean free path $\lambda_{free}$ for GaN at 60 keV ($\lambda_{free} \sim$ 120 nm) and for a t=80 nm thickness. Then, the average number of interactions is given by $\exp(-t/\lambda_{free}) \sim 0.5$ which gives about 1 inelastic interaction for every 2 electrons passing through the sample. Considering that most interactions are due to plasmon excitations and that a small number of electron-hole pairs (typically less than 3 [70,72]) are created for each plasmon decay, this gives a Fig. of about one electron-hole pair created per electron passing by. The value 7 to 100 electrons per photon is consistent considering that the emission rate is underestimated and different non-radiative channels are present. Moreover, the lower light emission from QDisk No. 19 and No. 16 in Fig. 10 is totally consistent with the lower energy shift observed in Fig. 7. In this case, non-radiative decay rate is sufficiently high to prevent carrier accumulation thus preventing the emission rate to be as high as in other QDisks. Such agreement gives confidence on the quantitative calibration of the emission efficiency scale in Fig. 10.

Fig. 10 also shows a smooth variation on the EQE with the excitation rate. We observed that up to $8*10^6$ electrons per second (for most QDisks), the emission efficiency increases remarkably, while QDisk No. 16 and No. 19 shows only a moderate increase. This efficiency increase at moderate excitation is indeed expected due to saturation of non-radiative defects, which scales linearly with the carrier density (the first term in the ABC model [9] for LEDs) [10] and also due to the increased overlap of electron and hole envelope functions for screened internal electric field. Hence, the radiative recombination rate increases and fewer carriers will be lost to trap-related non-radiative channels. Yet at higher electron beam currents, starting from $4*10^7$ electrons per second, the emission probability falls rather abruptly. Such a droop cannot be attributed to a sudden increase in the non-radiative paths due to irradiation damage, as it is a reversible process. We note that this analysis does not depend on the exact relation of emitted photons to detected counts on the CCD.

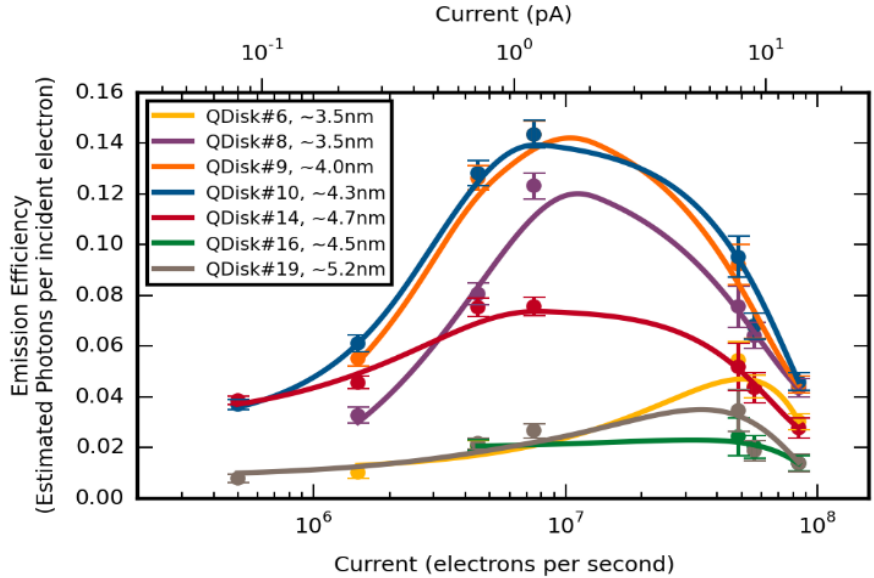

*Fig. 10: The number of photons detected per incident electron is plotted as a function of the exciting electron beam current for QDisks considered in Figure 7, from NW(b). The lines are guides to the eyes.*

While we are not aware of such a droop measurement in CL experiments, it is reminiscent of the quite extensively discussed droop in nitride LEDs [9]. Few possibilities are usually evoked to explain this droop in light emission devices [9]. These are defect assisted mechanisms, spontaneous emission reduction, Auger recombination and electron leakage. In a system without electric field and subsequent potential screening, all the effects have different dependences on the charge carrier's density that are usually expressed in the so-called "ABC" model [9]. In the absence of electron or charge-carrier leakage, the recombination is first dominated by non-radiative defects (low emission efficiency). This effect becomes negligible as the charge carrier density increases. So one can expect the emission efficiency to go to unity at high carrier density. We note that spontaneous emission rate can be reduced at high current, preventing the emission efficiency to go to unity, but it cannot account for a droop [9]. Now, in this scheme the Auger contribution becomes dominant at high charge carrier density and causes a droop of efficiency. In the present case of electrical field screening due to charge carrier density increase, the ABC model is much more difficult to apply, as the defect-related non-radiative rates and the radiative probabilities depend on the charge densities, because the screening induces a change in the electron and hole wavefunctions overlap. However, the induced changes in the probabilities are supposed to be the same in both situations and we can thus safely suppose that the screening itself cannot induce a droop. One probable cause for the droop is thus Auger effect, confirming recent findings in LEDs [12].

Interestingly, Fig.s 7 and 10 show that emission rates higher than $10^7$ photons per second cannot be reached in this system by an individual QDisk. Moreover, an excitation rate higher than 1 pA has a deleterious effect on the emission efficiency (Fig. 10). It is also noteworthy, that the obtained values for the maximum efficiency apply to the case of a high-energy electron excitation (60 keV acceleration voltage). Electrons with lower energy used in standard SEM-based CL set-ups transfer their energy to the sample much more efficiently. Therefore, at such low electron speed, lower currents should already lead to a high carrier density and result in a droop in emission efficiency. We note that a few picoamperes is a relatively low current for SEMs and hence for CL performed in SEM the probe current is likely to be greater than the value giving optimal efficiency.

# IV CONCLUSIONS

By using a nanometer-wide electron probe in a STEM, the light emission rate from 15 individual QDisks could be varied in a controlled way. The light emission rate is related to the carrier density generated by the electron beam inside each QDisk and hence could be controlled at will. Our observation validate a model in which, counterintuitively, a beam of high energy electrons, monitored with nanometre precision, can be ultimately seen as a controllable source of carriers generating roughly 1 carrier per incident electron.

We have shown that for thin QDisks, with emission energies greater than that of bulk GaN band gap, the emission energy is independent of the emission rate. This in because, since the emission energy in small QDisks is not significantly affected by the QCSE, screening the internal electric field has a minor effect. In general, thicker QDisks, with energies smaller than the bulk GaN, exhibit an emission energy that depends on the emission rate. Such dependence is explained as due to the QCSE and to the partial screening of the internal electric field by charge carriers inside the QDisk.

A semi-quantitative analysis of the emission efficiency as a function of the electron probe shows that the emission probability per incident electron increases up to a current of about 1 pA. This increase can be ascribed to the electron-hole envelope function overlap increase with carrier concentration and to the saturation of some non-radiative recombination paths. However, the emission probability per incident electron drops above a current of about 10pA. This efficiency droop can be tentatively attributed to Auger recombination within the QDisk.

# ACKNOWLEDGMENTS

This work has received support from CNPq and from projects 2014/23399-9 and 2012/10127-5, São Paulo Research Foundation (FAPESP). This work has received support from the French State managed by the National Agency for Research under the program of future investment EQUIPEX TEMPOS-CHROMATEM with the reference ANR-10-EQPX-50 as well as ANR "Investissement d'Avenir" programs "GaNeX" (ANR-11- LABX-2014) and "NanoSaclay" (ANR-10-LABX-0035). The research leading to these results has received funding from the European Union Seventh Framework Programme [#FP7/2007- 2013] under Grant Agreement #n312483 (ESTEEM2). The research described here has been partly supported by Triangle de la physique contract 2009-066T-eLight. TO and PAM acknowledge the support received from the European Union Seventh Framework Program under Grant Agreement 291 522-3DIMAGE. FDP and CD acknowledge funding from the ERC under grant number 25961 9 PHOTO EM.

# ANNEX

### EMISSION CALIBRATION RATE PROCEDURE

When needed, we have given an estimated emitted photon rate derived in the following way. Firstly, the loss due to all elements along the acquisition chain has been calculated based on their respective specifications. Secondly, we took into account the CCD camera quantum efficiency and sensitivity (counts per photon). Finally, to correct for the finite parabolic mirror collection efficiency, the deduced rate was divided by 0.36. This last estimation relies on the hypothesis that the objects under investigation are emitting isotropically. Considering all factors, we estimate that for each CCD count, approximately 30 (for Fig.s 3 (e) and (f) and Fig.s 7, 8 and 10) or 100 (Fig. 6) photons were emitted by the sample, the difference being due to different CCD cameras. Thus, this estimated rate, even if it is certainly just an approximation, should be proportional to the exact value. Anyway, the absolute value of the emission rate has no impact on the analysis and conclusions of the present paper. The whole procedure gave coherent results on data acquired with two different setups (except for the mirror which was the same) and several orientations of the NWs with respect to the axis of the parabolic mirror in a plane perpendicular to the electron beam.

### ASSIGNING SPECTRAL FEATURE TO QDISKS IN PRESENCE OF QCSE SCREENING

The emission obtained by exciting the QDisk No. 20 of NW(a), reported in Fig. 3 on the paper, is well spatially separated from the emission of all other QDisks, so it could be attributed to exclusively this specific QDisk. However, the situation is more complicated in the general case of partly spatially and spectrally overlapping emissions. To analyse the emission energy of individual QDisks when the emission energy possibly changes, we need to extend the analysis leading to the identification of individual QDisks luminescence described above and previously [40,41] to the case where the emission energy is shifting with the excitation rate (which can be varied by moving the excitation electron beam position). This is done by considering each spectrum individually and by considering regions of the SPIM in which the signal can be distinguished from the background. Such regions are determined by careful examination of the projected SPIM, which is represented as 2D images. This method was used for the analysis that leads to Fig. 11 .

In Fig. 11, the results for SPIMs on NWs (a), (c) and (d) are presented as 2D images, where one axis is the position of the beam along the NWs, the other axis is the emission energy, and the colour codes indicate the CL intensity. In relation to the CL data, the HAADF profile that was measured simultaneously is given, exhibiting a maximum at each QDisk position. One can see that very distinct features arise in the combined spatial-spectral maps. At energies higher than the bulk GaN band gap (see Fig. 11(b) and (e) – NW(b)QDisk No. 2), a diamond shaped emission pattern can be seen, similar to what was observed in [40] and [41], and the shape is centred on a given QDisk, as revealed by the HAADF profile. This supports the attribution of this feature to the emission of the corresponding QDisk. The diamond shape corresponds to constant emission energy. All other features, with energies below the GaN bandgap, have a triangular shape with the upper tip being brighter and centred on a specific QDisk (see Fig. 11 (a) and (d) – NW(a)QDisk No. 20). This shape is similar to what was observed by J. Lähnemann and co-workers when studying spontaneous polarization field in GaN.[73] We thus attribute each triangular shape to the specific QDisk on which it is centred. Note that the triangular shape is a synthetic representation of the effects seen on the filtered energy maps on Fig. 2.

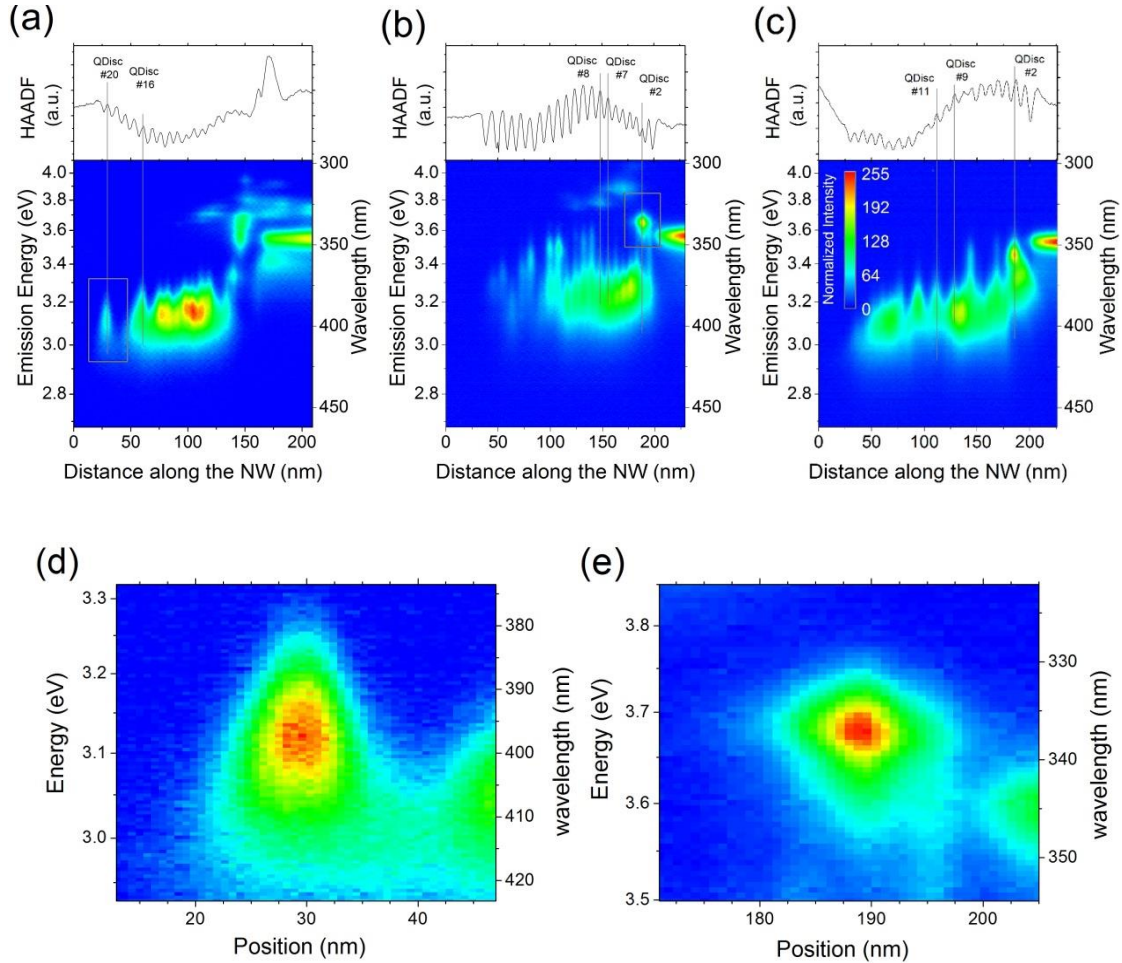

*Fig. 11: Spatial-spectral plots of three different nanowires containing 20 QDisks each. The position along the nanowire is indicated along the horizontal axis, while the vertical axis indicates the emission energy and the colour scale indicates the emission intensity. In (a), NW(a) is shown and similarly for NW(c) and NW(d), in (b) and (c). Fig.s (d) and (e) show details from the regions marked with rectangles in (a) and (b). The intense emission of the GaN segment covered with AlN is seen on the right hand side. Growth direction is from right to left.*

# REFERENCES

[1] N. P. Dasgupta , J. Sun , C. Liu , S. Brittman , S. C . Andrews, J. Lim , H. Gao , R. Yan , and Peidong Yang, Adv. Mater. **26**, 2137 (2014).

[2] C. M. Lieber, MRS Bulletin, **36**, 1052 (2011).

[3] Yat Li, Fang Qian, Jie Xiang, and Charles M. Lieber Mater. Today **9**, 18 (2006).

[4] M.J. Holmes, K. Choi, S. Kako, M. Arita, and Y. Arakawa, Nano Lett. **14**, pp 982-986 (2014).

[5] L. Rigutti et al., NanoLetter **10**, pp.2939-2943 (2010).

[6] M. Tchernycheva et al, NanoLetters **14**, pp. 2456-2465 (2014).

[7] S. Li and A. Waag, J. Appl. Phys. **111**, 071101 (2012).

[8] H.P.T. Nguyen, S. Zhang, K. Cui, X. Han, S. Fathololoumi, M. Couillard, G. A. Botton, and Z. Mi, Nano Lett. **11**, 1919 (2011).

[9] J. Piprek, Phys. Status Solidi A **207**, 2217–2225 (2010).

[10] Q. Dai *et al.*, Appl. Phys. Lett., **97**, 133507 (2010).

[11] T. Kohno *et al.*, Jpn. J. Appl. Phys. **51**, 072102 (2012).

[12] J. Iveland, L. Martinelli, J. Peretti, J. S. Speck, C. Weisbuch, Phys. Rev. Lett. **110**, 177406 (2013).

[13] H. Murotani *et al.*, Jpn. J. Appl. Phys. **52**, 08JE10 (2013).

[14] W. Guo, M. Zhang, P. Bhattacharya, and J. Heo Nano Lett. **11**, 1434–1438 (2011).

[15] M. Peter, A. Laubsch, W. Bergbauer, T. Meyer, M. Sabathil, J.Baur, and B. Hahn, Phys. Status Solidi A **206**, 1125 (2009).

[16] F. Bernardini, V. Fiorentini, and D. Vanderbilt, Phys. Rev. B, **56**, pp. R10024–R10027 (1997).

[17] M. Leroux, N. Grandjean, M. Laugt, J. Massies, B. Gil, P. Lefebvre, and P. Bigenwald. Phys. Rev. B, **58**, pp. R13371–R13374, (1998).

[18] D. A. B. Miller, D. S. Chemla, T. C. Damen, A. C. Gossard, W. Wiegmann, T. H. Wood, and C. A. Burrus, Phys. Rev. Lett. **53**, 2173 (1984).

[19] P. Lefebvre and B. Gayral, C. R. Physique **9**, 816 (2008).

[20] P. Bigenwald, A. Kavokin, B. Gil, and P. Lefebvre, Phys. Rev. B, **61**, 15621 (2000).

[21] S. Kalliakos, T. Bretagnon, P. Lefebvre, T. Taliercio, B. Gil, N. Grandjean, B. Damilano, A. Dussaigne, and J. Massies, J. Appl. Phys. **96**, 180 (2004).

[22] Widmann, J. Simon, B. Daudin, G. Feuillet, J. L. Rouvière, N. T. Pelekanos and G. Fishman, Phys. Rev. B **58**, R15989 (1998).

[23] T. Bretagnon, P. Lefebvre, P. Valvin, R. Bardoux, T. Guillet, T. Taliercio, B. Gil, N. Grandjean, F. Semond, B. Damilano, A. Dussaigne, and J. Massies Phys. Rev. B **73**, 113304 (2006).

[24] S. Kalliakos, P. Lefebvre, and T. Taliercio, Phys. Rev. B **67**, 205307 (2003).

[25] A. Bell, *et al.*, Appl. Phys. Lett. **84**, 58 (2004).

[26] S. Sanguinetti, M. Gurioli, E. Grilli, M. Guzzi, and M. Henini, Appl. Phys. Lett. 77, 1982 (2000).

[27] J.H. Ryou, W. Lee, J. Limb, D. Yoo, J.P. Liu, R.D. Dupuis, Z.H. Wu, A.M. Fischer, and F.A. Ponce, Appl. Phys. Lett. **92**, 101113 (2008).

[28] M. Gurioli, S. Sanguinetti, and M. Henini, Appl. Phys. Lett. **78**, 931 (2001).

[29] J. Renard, R. Songmuang, G. Tourbot, C. Bougerol, B. Daudin, and B. Gayral, Phys. Rev. B **80**, 121305(R) (2009).

[30] S.A. Empedocles and M. G. Bawendi, Science **278**, 2114 (1997).

[31] O. Ambacher, J. Majewski, C. Miskys, A. Link, M. Hermann, M. Eickhoff, M. Stutzmann, F. Bernardini, V. Fiorentini, V. Tilak, B. Schaff, and L. Eastman, J. Phys.: Condens. Matter **14**, 3399 (2002).

[32] T. Kuroda and A. Tackeuchi, J. Appl. Phys. **92**, 3071 (2002).

[33] C. Netzel, V. Hoffmann, T. Wernicke, A. Knauer, M. Weyers, M. Kneissl, and N. Szabo, Journal of Applied Physics **107**, 033510 (2010)

[34] Jong-In Shim, Hyunsung Kima, Dong-Soo Shinb, and Han-Youl Ryu, Proc. of SPIE Vol. **7939**, 79391A-1 (2011);

[35] J. Bruckbauer, P. R Edwards, J. Bai, T. Wang and R. W Martin, Nanotechnology **24** (2013) 365704

[36] James R. Riley, Sonal Padalkar, Qiming Li, Ping Lu, Daniel D. Koleske, Jonathan J. Wierer, George T. Wang, and Lincoln J. Lauhon, Nano Lett. **13**, 4317–4325 (2013).

[37] Rigutti *et al.*, Nanoletters **14**, 107–114 (2014).

[38] Yong-Ho Ra, Rangaswamy Navamathavan, Hee-Il Yoo, and Cheul-Ro Lee, Nano Letters **14**, 1537-1545 (2013).

[39] Hieu Pham Trung Nguyen, Mehrdad Djavid, Kai Cui and Zetian Mi Nanotechnology **23**, 194012 (2012).

[40]   L.F. Zagonel, S. Mazzucco, M. Tencé, K. March, R. Bernard, B. Laslier, G. Jacopin, M. Tchernycheva, L. Rigutti, F.H. Julien, R. Songmuang, and M. Kociak, Nano Lett. **11**, 568 (2011).

[41]   L.F. Zagonel, L. Rigutti, M. Tchernycheva, G. Jacopin, R. Songmuang, and M. Kociak, Nanotechnology 23, 455205 (2012).

[42]   M. Kociak, O. Stéphan, A. Gloter, L.F. Zagonel, L.H.G. Tizei, M. Tencé, K. March, J.D. Blazit, Z. Mahfoud, A. Losquin, S. Meuret, and C. Colliex, Comptes Rendus Physique **15**, 158 (2014).

[43]   M. A. Herman, D. Bimberg, and J. Christen, Journal of Applied Physics 70, R1-R52 (1991).

[44]   J. Barjon, J. Brault, B. Daudin, D. Jalabert and B. Sieber, Journal of Applied Physics **94**, 2755 (2003).

[45]   U. Jahn, J. Ristić, and E. Calleja, Appl. Phys. Lett. **90**, 161117 (2007).

[46]   P.R. Edwards, L.K. Jagadamma, J. Bruckbauer, C. Liu, P. Shields, D. Allsopp, T. Wang, and R.W. Martin, Microscopy and Microanalysis **18**, 1212 (2012).

[47]   M. Albrecht, L. Lymperakis, J. Neugebauer, J. Northrup, L. Kirste, M. Leroux, I. Grzegory, S. Porowski, and H. P. Strunk, Phys. Rev. B **71**, 035314 (2005).

[48]   A. Pierret, C. Bougerol, B. Gayral, M. Kociak, and B. Daudin, Nanotechnology **24**, 305703 (2013).

[49]   M. Albrecht, H. Strunk, J. Weyher, I. Grzegory, S. Porowski and T. Wosinski, J. Appl. Phys. **92**, 2000 (2002).

[50]   S.K. Lim, M. Brewster, F. Qian, Y. Li, C.M. Lieber, and S. Gradečak, Nano Lett. **9**, 3940 (2009).

[51]   M. Grundmann *et al.*, Phys. Rev. Lett. **74**, 4043 (1995).

[52]   L.H.G. Tizei and M. Kociak, Nanotechnology **23**, 175702 (2012).

[53]   L.H.G. Tizei and M. Kociak, Phys. Rev. Lett. **110**, 153604 (2013).

[54]   J. Barjon, P. Desfonds, M.A. Pinault, T. Kociniewski, F. Jomard, and J. Chevallier, Journal of Applied Physics **101**, 113701 (2007).

[55]   E.J.R. Vesseur, R. de Waele, M. Kuttge, and A. Polman, Nano Lett. **7**, 2843 (2007).

[56]   P.R. Edwards, D. Sleith, A.W. Wark, and R.W. Martin, J. Phys. Chem. C 115, 14031 (2011).

[57]   Arthur Losquin, Luiz F. Zagonel, Viktor Myroshnychenko, Benito Rodríguez-González, Marcel Tencé, Leonardo Scarabelli, Jens Förstner, Luis M. Liz-Marzán, F.

Javier García de Abajo, Odile Stéphan, and Mathieu Kociak, *Nano Lett.*, 15 (2), pp 1229–1237 (2015).

[58] P. Das, T.K. Chini, and J. Pond, J. Phys. Chem. C 116, 15610 (2012).

[59] G. Tourbot, C. Bougerol, F. Glas, L. F. Zagonel, Z. Mahfoud, S. Meuret, P. Gilet, et al. Nanotechnology **23** (13), 135703.

[60] R. Songmuang, O. Landré and B. Daudin, Appl. Phys. Lett. **91**, 215902 (2007).

[61] M. Tchernycheva *et al.*, Nanotechnology **18,** 385306 (2007).

[62] de la Peña, F. , Burdet, P., Sarahan, M., Nord, M., Ostasevicius, T., Taillon, J., Eljarrat, A., Mazzucco, S., Fauske, V.T., Donval, G., Zagonel, L.F., Walls, M., Iyengar, I., HyperSpy 0.8. d.o.i.: 10.5281/zenodo.16850

[63] See Supplemental Material at [URL will be inserted by publisher] for video with the full Spectral Image in a broad spectral range.

[64] J. Lähnemann, U. Jahn, O. Brandt, T. Flissikowski, P. Dogan, and H. T. Grahn, J. Phys. D: Appl. Phys. **47**, 423001 (2014).

[65] T. Kuroda, A. Tackeuchi, and T. Sota, Appl. Phys. Lett. **76**, 3753 (2000).

[66] E. Kuokstis, C.Q. Chen, M.E. Gaevski, W.H. Sun, J.W. Yang, G. Simin, M. Asif Khan, H.P. Maruska, D.W. Hill, M.C. Chou, J.J. Gallagher, and B. Chai, Appl. Phys. Lett. **81**, 4130 (2002).

[67] A. Chernikov, S. Schäfer, M. Koch, S. Chatterjee, B. Laumer, and M. Eickhoff, Phys. Rev. B **87**, 035309 (2013).

[68] R. Leavitt, J. Little, Physical Review B, 42(18), 11774–11783 (1990).

[69] E. Kioupakis, Q. Yan and C. G. Van de Walle Appl. Phys. Lett. 101, 231107 (2012).

[70] S. Meuret, et al. Phys. Rev. Lett. **114**, 197401 (2015).

[71] G. Rossbach, J. Levrat, G. Jacopin, M. Shahmohammadi, J.-F. Carlin, J.-D. Ganière, R. Butté, B. Deveaud, and N. Grandjean, Phys. Rev. B, **90**, 201308(R), 2014.

[72] A. Rothwarf, Journal of Applied Physics **44**, 752 (1973).

[73] J. Lähnemann et al. Phys. Rev. B **86** 081302 (2012).